\documentclass[fleqn,10pt]{SelfArx} 

\definecolor{color1}{RGB}{0,0,90} 
\definecolor{color2}{RGB}{0,20,20} 

\newlength{\tocsep} 
\usepackage{lipsum} 

\usepackage{url}

\JournalInfo{Palo Alto Research Center} 
\Archive{} 

\PaperTitle{CCNx 1.0 Bidirectional Streams} 

\Authors{Marc Mosko\textsuperscript{1}*} 
\affiliation{\textsuperscript{1}\textit{Palo Alto Research Center}} 
\affiliation{*\textbf{Corresponding author}: marc.mosko@parc.com} 

\newcommand{\keywordname}{Keywords} 

\newcommand{\parccopyright}{} 

\fancypagestyle{empty}{ %
  \fancyhf{} 
  \fancyfoot[EOC]{\parccopyright}
}

\Abstract{In Content Centric Networks, where one retrieves data based on a given name, not a conventional connection
to a server or other device, there is a need for a standard mechanism to establish bi-directional streams.  We describe
a method to open and maintain such conversations using the techniques of a named data network.}

\begin{document}

\flushbottom 

\maketitle 

\tableofcontents 

\thispagestyle{empty} 


\section*{Introduction} 

\addcontentsline{toc}{section}{\hspace*{-\tocsep}Introduction} 

CCNx Bidirectional Streams describes how two parties may open
bidirectional data streams using networking techniques of CCNx.
In CCNx, prates exchange data based on descriptive names of the data,
not on machine addresses.  This allows for more diverse forms of networking than
traditional TCP/IP.  In some cases, however, parties may still wish to directly
exchange byte streams between two ``live'' applications and thus need
the techniques described here.

Bidirectional Streams describes a Basic Setup and Authenticated Setup.  In
Basic Setup, the parties do not authenticate each other upfront, but defer
it to later optional verification of the exchanged data.  In Authenticated
Setup, both parties authenticate each other before exchanging data,
and they may optionally use data encryption.

Section~\ref{sec:ccnx} describes the Content Centric Networking (CCN)
approach to Named Data and previous work on Voice over CCN that solved
a specific type of bi-directional data.  Section~\ref{sec:bidir} presents
this paper's contribution on Bidirectional Streams and describes both
basic setup and authenticated setup.  Section~\ref{sec:forward}
describes how Bidirectional Streams works when an initial setup request
does not reach the best site to service the stream, and the request
is forwarded to a different site.


\section{CCNx Overview}\label{sec:ccnx}

In a Content Centric Network (CCN)~\cite{conext09ccn}, devices exchange data on a network based on
descriptive names for the data, not on machine addresses such as in Internet Protocol.  
In summary, CCN is based on the exchange of Interest messages asking for content and
Content Objects that contain data.
A device
requests content by issuing an Interest for the data using a name, such as "ccnx:/newyorktimes/frontpage".
The Interest travels through the network until a device has a corresponding Content Object with a matching
name.  The matching Content Object travels on the Interest's reverse path and provides the data
to the original device. 

The Interest/Content Object exchange is a uni-directional flow of data.  The Interest contains a name prefix
that enables the message to be routed over a network and usually contains information needed to select
a specific content object or initiate a specific service at the Interest consumer.  Application data then travels
back in response to the Interest using a similar name to the Interest.  What is missing is a conventional way
to establish bi-directional data exchanges using simple names.

Bidirectional streams over Named Networks extends the methods originally proposed for
Voice over CCN (VoCCN)~\cite{voccn-rearch}.  VoCCN describes a method to initiate a Voice-over-IP
conversation by exchanging SIP messages in Interests and Content Objects, then using
bi-directional named streams for the Real Time Protocol (RTP) packet exchange.  We generalize
this method for arbitrary data connections between devices.

\subsection{Voice over CCN}
In Voice over CCN (VoCCN), a caller sends a specially formed Interest to the callee.
The Interest contains the routable name of the callee and also encodes a SIP Invite~\cite{sip}
and a symmetric encryption key.  The symmetric encryption key is encrypted with the callee's 
public key, so only they can decrypt it.  The SIP Invite is encrypted with the symmetric key.
The callee inspects the SIP Invite and extracts the caller's identity, which includes their
routable name and the call-id.  The callee creates a response Content Object that
includes the callee's SIP Accept.
Based on these exchanges over CCN, the caller and callee can create a bi-directional
RTP~\cite{rtp} flow, that is optionally encrypted with the symmetric key.


\section{Bidirectional Streams}\label{sec:bidir}

To establish bi-directional streams, at least one party must register to receive 
Interests on a service name for \textit{service rendezvous}.  The
initiator sends to the service name an Interest that specifies the
initiator's stream name.  The provider responds with a Content Object
that contains the provider's stream name.  The parties may now exchange
data over the two streams using segmented Content Objects.
When one party is done writing to its stream, it indicates the final
segment by setting the FinalBlockID of the ContentObject to that final
segment number.  If no final data is necessary, sets the Content size
to zero.  When the second party is done
writing to its stream, it closes it with the same technique.

\begin{figure}[ht]\centering
\includegraphics[page=1,width=\linewidth,trim=.25in .25in 3in .25in]{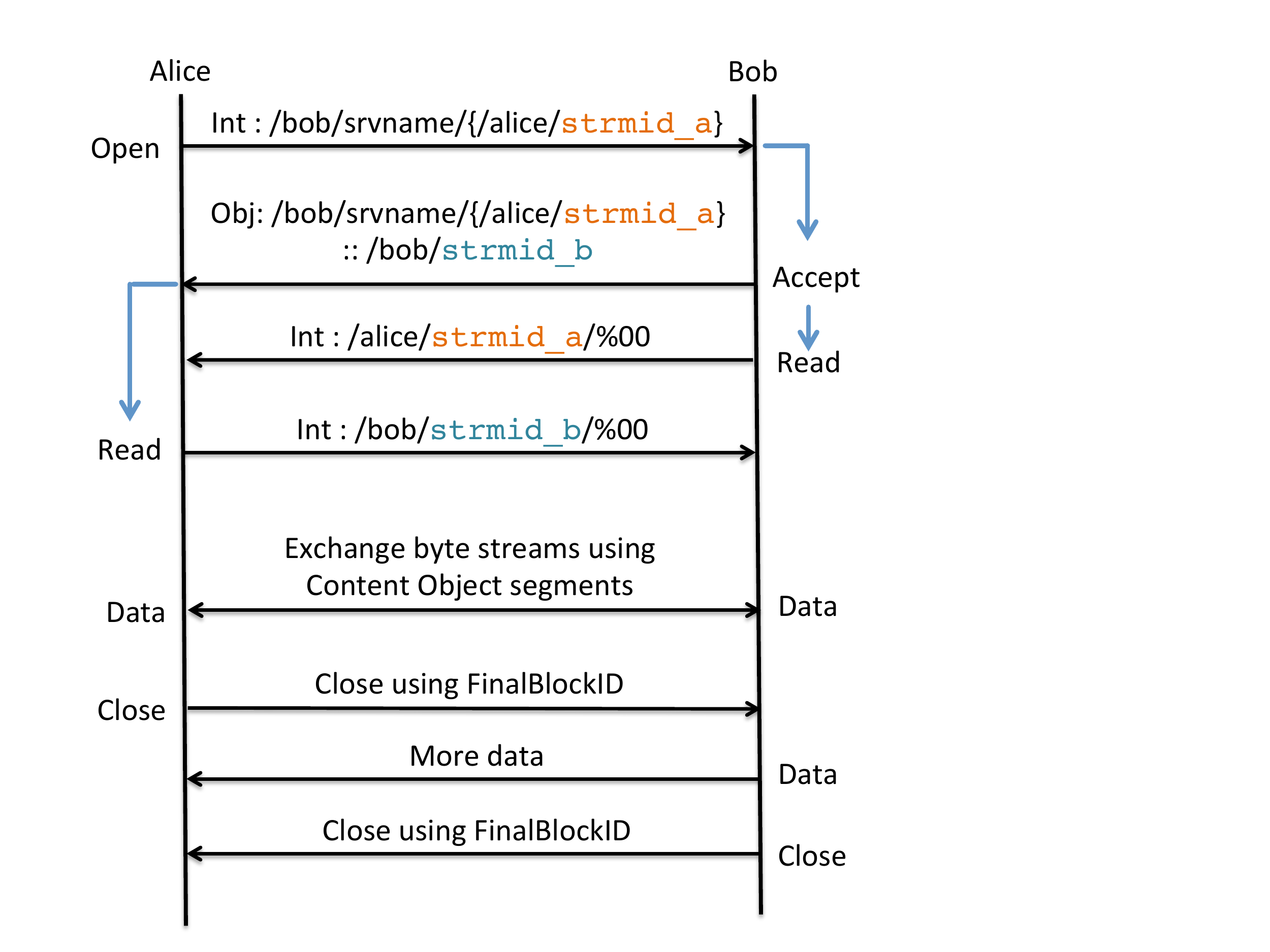}
\caption{Basic Stream Setup}
\label{fig:basicsetup}
\end{figure}

We use the notation \texttt{Int: <name>} to indicate an Interest
with the name \texttt{<name>}.  Other CCN selectors may
appear to facilitate reading a CCN stream, but for brevity we do not
detail them here.  We use the notation \texttt{Obj: <name> :: <contents>}
to represent a Content Object with the name \texttt{<name>}
and its payload contents \texttt{<contents>}.  We assume
the contents are encoded using a standard method, such as
XML objects or CCNB encoding.  The actual stream payloads,
noted as ``Data'' exchanges in Fig.~\ref{fig:basicsetup}, have
arbitrary user bytes as payload and are not interpreted by 
the bidirectional protocol.

Fig.~\ref{fig:basicsetup} depicts the basic setup mechanism.  
Bob has registered to receive interests for the name \url{/bob/srvname}.
Alice sends Bob an Interest where the last name component encodes
Alice's stream name \url{/alice/strmid_a}, where \url{strmid_a}
is a locally unique random number.  Bob responds with an Accept
Content Object.  The object's name exactly matches the Interest's name
and contains the Name of Bob's stream name, where \url{strmid_b}
is a locally unique random number.  Bob also begins expressing
Interests for the first segment of Alice's stream.  Once Alice
receives the Accept object, she begins expressing Interests for
Bob's stream.  The two parties now exchange data over segmented
Content Objects. 

Bidirectional stream readers should use flow control and re-express Interests
for missing segments.  The particulars of flow control and retransmissions
are beyond the scope of this document.

In Basic setup, Alice may assure herself that the Accept object came from
Bob, because like all CCN Content Objects, it is signed by Bob.

One problem with basic stream setup is that Bob has no assurance
that Alice is sending him a valid stream name.  It could be the
impostor Eve trying to DDoS Bob by causing him to generate
Accept objects, or DDoS Alice by having Bob flood Alice with Read Interests.
In Section~\ref{sec:trusted} we discuss using authenticated setup. 

\subsection{Closing a stream}
A Content Object contains an optional field called the FinalBlockID.  To close a stream,
the writer sends a Content Object where the FinalBlockID is set to the object's segment
number.  If no data needs to be sent, the object's Content block has zero length.

When a stream reader receives a final block, it should cancel any pending interests
for segments beyond the final block.

\subsection{Authenticated streams}\label{sec:trusted}
We saw how basic stream setup is susceptible to denial-of-service attacks
because Bob does not know Alice's identity during the Accept phase.
This could cause Bob to send unwanted Read Interests to some third
party.  In this section, we present an authenticated stream setup.

We use the notation $S_A{data}$ to mean that $data$ is signed by the
private key of $A$ so someone with the public key of $A$ could verify the
signature.  The notation $E_B{data}$ means that the data is encrypted
with the public key of $B$ so someone with the private key of $B$
could decrypt it.

\begin{figure}[ht]\centering
\includegraphics[page=2,width=\linewidth,trim=.25in 4.25in 3in .25in]{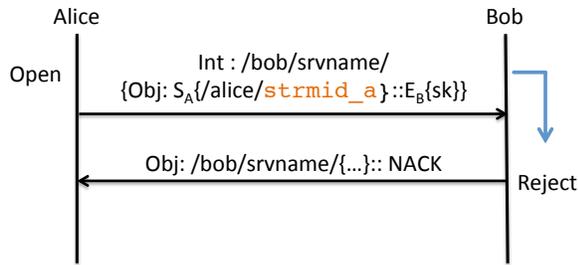}
\caption{Authenticated Stream Setup (showing Reject)}
\label{fig:authsetup}
\end{figure}

Fig.~\ref{fig:authsetup} shows an authenticated stream setup.  The
initial Interest that Alice sends to Bob contains a signed Content Object
along with a symmetric encryption key.  The Content Object name is
Alice's stream name, and it is signed with Alice's key.  Bob may now
apply a trust model to determine if Alice is a valid user, and to determine
if Alice is authorized to use the namespace of her stream name.
If Bob does not trust the provided information, he sends a Reject
Content Object along the Interest reverse path.

\begin{figure}[ht]\centering
\includegraphics[page=3,width=\linewidth,trim=.25in .25in 3in .25in]{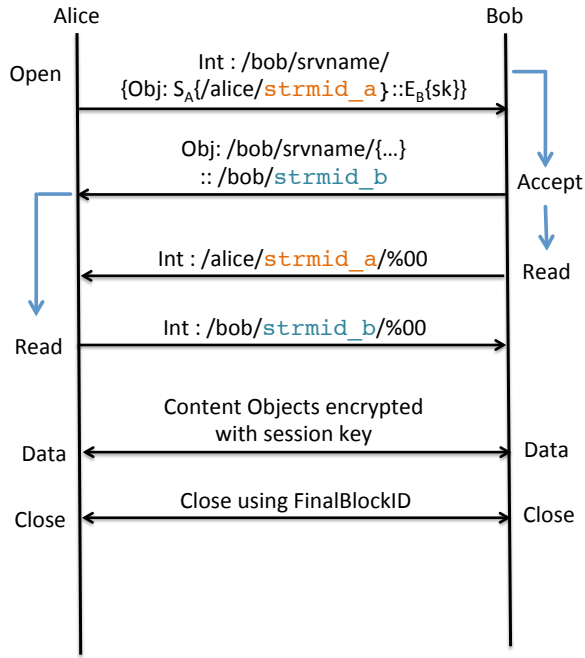}
\caption{Authenticated Stream Setup (showing Accept)}
\label{fig:authaccept}
\end{figure}

Fig.~\ref{fig:authaccept} shows an example of Bob accepting Alice's 
authenticated setup.  The Accept Content Object's name must exactly
match Alice's Interest name, and it contains Bob's stream name.

\subsection{Encryption}
When Alice uses Authenticated Stream Setup, she may include an 
optional symmetric session key if she plans to encrypt the contents
of her Content Objects during Data exchange.  If she does not
plan to use encryption, she may omit the session key.

If Bob plans to encrypt his Data content, he may use Alice's
session key, if provided.  Otherwise, he must include his own
session key in his Accept response.

The protocol does not encrypt the stream names when exchanged in
authenticated Open interests and Accept objects.  That is because
those names will be plaintext in the ensuring data conversation.

All Content Objects are signed, as normal, in CCN.  Alice and Bob
may verify each object as it is received.


\section{Anycast and Redirection}\label{sec:forward}
In another scenario, the service name \url{ccnx:/bob/srvname}
may be used like an Anycast address, getting Alice to a nearby
service provider.  That provider, however, may not be the best
to service Alice.  In this case, the first site, \url{bob_1} forwards
Alice's Open interest to site \url{bob_2}, and data then flows
directly between Alice and \url{bob_2}.

\begin{figure}[ht]\centering
\includegraphics[page=4,width=\linewidth,trim=.25in 3.5in 3in .25in]{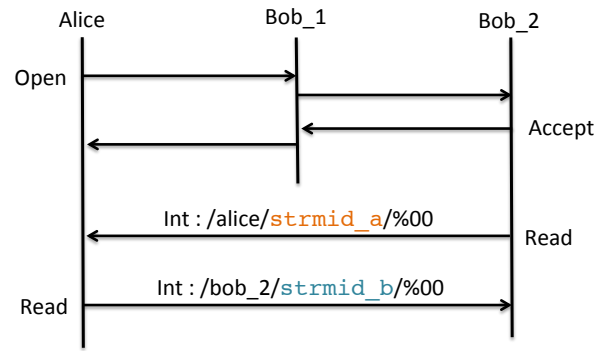}
\caption{Connection forwarding}
\label{fig:forward}
\end{figure}

Fig.~\ref{fig:forward} depicts this scenario.  Forwarding
applies to both Basic Setup and Authenticated Setup.
The Open request is forwarded from \url{bob_1} to
\url{bob_2}, and the Accept object follows the reverse
path of the Interests.  The Accept content object
carries \url{bob_2}'s stream name, so in the Data Exchange
phase, Alice and \url{bob_2} communicate directly
without \url{bob_1}.

In Basic Setup, the second provider does not necessarily need
to be related to the first provider.  For example, Bob could
forward to Carol, so long as Alice is willing to accept an 
Accept Content Object signed by Carol.  In Authenticated
Setup, it may still be possible to use a 3rd party for the
second service provider, but one would need to assure
that Alice's trust model allows for it, and that the second
service provider can decrypt the session key, if Alice
included one.

\section{Conclusion}
Bidirectional Streams over Named Networks describes a general purpose framework
for two parties to open bidirectional streams using Named networking conventions.
Basic setup does not use client authentication or encryption.  Authenticated
Setup allows both parties to authenticate each other and allows for optional
encrypting using a session key.



\bibliographystyle{unsrt}
\bibliography{bidir}


\end{document}